\documentclass[useAMS,usenatbib,usegraphicx]{mn2e}
\bibpunct[, ]{}{}{;}{a}{,}{,}

\newcommand{\bl}[1]{\mbox{\boldmath$ #1 $}}

\begin{document}

\title[Flattening of abundance gradients in spiral galaxies]{Spiral stellar density waves and the flattening of abundance
gradients in the warm gas component of spiral galaxies.}
\author[E. I. Vorobyov]{E. I. Vorobyov
\thanks{E-mail:vorobyov@astro.uwo.ca (EIV)} \\
$^{1}$CITA National Fellow, Department of Physics and Astronomy, University of Western Ontario, 
London, Ontario, N6A 3K7, Canada.}

\maketitle

\begin{abstract}{Motivated by recent observations of plateaus and minima
in the radial abundance distributions of heavy elements in the Milky Way and
some other spiral galaxies, we propose a dynamical mechanism for the
formation of such features around corotation. Our numerical simulations show 
that the non-axisymmetric gravitational field of spiral density waves generates 
cyclone and anticylone gas flows in the vicinity of corotation. 
The anticyclones flatten the  pre-existing negative
abundance gradients by exporting many more atoms of heavy elements outside
corotation than importing inside it. This process is very efficient and
forms plateaus of several kiloparsec in size around corotation 
after two revolution periods of a galaxy. The strength of anticyclones
and, consequently, the sizes of plateaus depend on the pitch angle of spiral arms and 
are expected to increase along the Hubble sequence.}
\end{abstract}


\section{Introduction}

It has recently been recognized that density waves in the stellar component
of spiral galaxies have a profound effect on the dynamics of stars, cold
gas clouds, and dust in the vicinity of the corotation resonance. The non-axisymmetric gravitational
field of spiral stellar density waves causes large changes ($\sim 50$ per cent
over the lifetime of a galaxy) in the angular momenta of individual stars
and cold gas clouds around the corotation radius (Sellwood \& Binney \cite{Sellwood}). 
Considerable radial migrations associated with the angular momentum changes
are expected to dilute the abundance gradients in the cold gas component of
spiral galaxies (Sellwood \& Preto \cite{Sellwood2}).

The radial abundance distribution of heavy elements in log scale in at
least some spiral galaxies
(perhaps, including the Milky Way) cannot be described by a linear function with a negative slope. 
According to Zaritsky, Kennicutt, \& Huchra (\cite{Zaritsky}), the oxygen abundances 
in NGC~2997, NGC~3319, NGC~5033 and other spiral galaxies in their sample show 
a complex nonlinear behaviour -- plateaus and minima can
be identified in the radial distribution of oxygen. 
Perhaps more convincing evidence for a complex radial distribution of heavy
element abundances is found in the Milky Way. For instance, the radial abundance
distributions of O, N, Mg, and other heavy elements derived by Daflon \& Cunha (\cite{Daflon})
from a sample of OB stars show a minimum near 8~kpc (although they have
not accentuated the importance of this behaviour and approximated
the radial abundance profiles by a linear function with a negative slope).  The existence
of a plateau in the oxygen abundance distribution has also been reported
by Andrievsky et al. (\cite{Andrievsky}).
Although a definite confirmation of plateaus or/and minima in the radial
abundance profiles of heavy elements requires a larger sample of abundance
tracers than has been used in the abovementioned studies, the existing evidence
strongly suggests these features.

A simple multizone model of chemical enrichment in spiral galaxies has been recently proposed 
by Mishurov et al. (\cite{Mishurov})
and Acharova et al. (\cite{Acharova}). It explains the formation of minima and/or plateaus
in the radial abundances of heavy elements near corotation by a selective
action of star formation. The star formation rate around corotation is
assumed to have a minimum (due to the lack of strong spiral shock waves) and
consequently the heavy element production also has a minimum at the corotation
radius. The assumed temporal migration of the corotation resonance can produce
either plateaus or minima in the radial abundance distributions of heavy
elements.
 
In this paper, we focus on a purely hydrodynamic explanation for a 
nonlinear radial distribution of heavy elements in spiral galaxies.
We present the first numerical hydrodynamic simulations
that self-consistently explain the formation of a plateau in the heavy
element abundance distribution in the vicinity of corotation.  
We demonstrate the development of cyclones and anticyclones
in the gas flow around corotation and study their influence on
the radial abundance distribution of heavy elements in spiral galaxies.
The existence of cyclones and anticyclones  has
been observationally confirmed in at least two spiral galaxies (Fridman
et al. \cite{Fridman1,Fridman2}) and has been predicted in the laboratory
experiments of rotating shallow water modelling (Nezlin \cite{Nezlin}). 
The model equations are formulated in \S~\ref{model}. The numerical code
is described in \S~\ref{code} and the initial conditions are given in \S~\ref{init}.
The results of numerical simulations are presented in \S~\ref{results}.
The possible implications for spiral galaxies of different Hubble types
are discussed in \S~\ref{hubble}. The main results are summarized in \S~\ref{sum}.

\section{Model description and basic equations}
\label{model}
Our model galaxy consists of a gas disk that evolves in the external gravitational potential 
of the spherical dark matter halo and stellar disk. We study the behaviour of the warm 
gas component ($T\sim 10^4$~K) and the heavy element admixtures within it, which can be adequately 
described by the equations of hydrodynamics.
The stellar disk is split into axisymmetric and 
non-axisymmetric components. The axisymmetric component is assumed to have a power-law
radial density profile, whereas the non-axisymmetric component is described by 
a running spiral density wave (Lin et al. \cite{Lin}). The density profile of the spherical dark matter halo is assumed to 
be that of a modified isothermal sphere (Binney \& Tremaine \cite{BT}).

We use the physically motivated thin-disk approximation to
describe the motion of gas and heavy elements in the external gravitational potential of the 
dark matter halo and stellar disk. In this approximation, the radial extent of 
the gas disk is assumed to be much larger than its vertical height, discarding the need
to solve for the vertical motion of gas.
We assume that heavy elements are collisionally coupled to
the gas, which eliminates the need to solve the equations of motion for
the heavy elements.
The basic equations governing the dynamics of the gas and heavy element components
are

\begin{equation}
{\partial \Sigma_{\rm x} \over \partial t} = - {\bl \nabla} \cdot ({\bl v} \Sigma_{\rm x}),
\label{first}
\end{equation}

\begin{equation}
{\partial \Sigma_{\rm g} \over \partial t} = - {\bl \nabla} \cdot ({\bl v} \Sigma_{\rm g}),
\end{equation}

\begin{equation}
{\partial {\bl v} \over \partial t} +({\bl v} \cdot {\bl
\nabla}) {\bl v} = -{\bl \nabla} \Phi_{\rm s1,s2,h} - {{\bl \nabla} P
 \over \Sigma_{\rm g}}.
 \label{third}
\end{equation}
Here, $\Sigma_{\rm g}$ and $\Sigma_{\rm x}$ are the surface densities of
the gas and heavy element components, respectively, $\bl v$ is the gas velocity
in the disk plane, and $P=c_{s}^2 \Sigma$ is the vertically integrated gas pressure.
The gas disk is assumed to be isothermal at $T=10^4$~K, which yields a value
of $c_{\rm s}=9.12$~km~s$^{-1}$ for the isothermal sound speed. 
We note that if the ratio $\Sigma_{\rm x}/\Sigma_{\rm g}$ is constant in
the disk at any given time, it will continue to be constant at all times,
because $\Sigma_{\rm x}$ will be advected exactly in the same manner as
$\Sigma_{\rm g}$. The expressions
for the gravity force due to the gravitational potentials $\Phi_{\rm h}$,
$\Phi_{\rm s1}$, and $\Phi_{\rm s2}$ of the dark matter halo, axisymmetric
and non-axisymmetric stellar components, respectively, are given below.

The spherical dark matter halo has a density profile described by 
the modified isothermal sphere
\begin{equation}
\rho_{\rm h}={\rho_{\rm h0}\over (1+r/r_{\rm h})^2},
\label{halodens}
\end{equation}
where $\rho_{\rm h0}$ and $r_{\rm h}$  are the central volume density and characteristic
scale length of the dark matter halo, the values of which are fixed by the halo mass $M_{\rm h}$
as described in Vorobyov \& Shchekinov  (\cite{VS}) and references therein.  In the following text we assume $M_{\rm
h}=10^{12}~M_{\odot}$. The radial gravity force of the spherical dark matter halo in the plane of the
gas disk can be written as 
\begin{equation}
{\partial \Phi_{\rm h} \over \partial r}=4 \pi G \rho_{\rm h0}
r_{\rm h}\left[ r/r_{\rm h} - \arctan(r/r_{\rm h}) \right]
\left({r_{\rm h}\over r}\right)^2.
\label{halo}
\end{equation}

The axisymmetric component of the stellar disk is assumed to have a power-law radial density profile
of the form
\begin{equation}
\Sigma_{\rm s}(r)= {B^2 \over 2 \pi G} \left[(r_{\rm s}^2+r^2)^{-3/2} \right],
\label{stellar}
\end{equation}
where $B^2= 2 \pi G r_{\rm s}^3 \Sigma_{\rm s0}$.
The radial component of the gravity force of such a density distribution
is given by Toomre (\cite{Toomre})
\begin{equation}
{\partial \Phi_{\rm s1} \over \partial r}=B^2\left[{r\over r_{\rm s}} (r_{\rm s}^2+r^2)^{-3/2}\right].
\label{sym}
\end{equation}
In the following, we use the central stellar density
$\Sigma_{\rm s0}=1.2\times 10^3~M_\odot$~pc$^{-2}$ and $r_{\rm s}=3$~kpc,
which gives us a total stellar mass of $M_{\rm st}=7\times 10^{10}~M_\odot$.

The non-axisymmetric part of the stellar gravitational potential is described in
the polar coordinates ($r,\phi$) by a running
density wave as (Lin et al. \cite{Lin}, L{\'e}pine et al.  \cite{Lepine}, Vorobyov
\& Shchekinov \cite{VS})
\begin{equation}
\Phi_{\rm s2}(r,\phi)=-C(r) \cos\left[ m(\cot(i) \ln(r/r_{\rm sp})+\phi - \Omega_{\rm sp} t)\right],
\label{nonsym}
\end{equation}
where $C(r)$ is the radially varying amplitude of the stellar gravitational
potential, $i$ is the pitch angle, $r_{\rm sp}$ is the characteristic radius of the spiral at $\phi=0$ , 
$m$ is the number of spiral arms, and $\Omega_{\rm sp}$ is the
angular velocity of spiral pattern.
In the following we adopt $m=2$, $r_{\rm sp}=6$~kpc and $\Omega_{\rm sp}=21$~km~s$^{-1}$~kpc. 
The value of $i$ is varied to study its influence on the radial distribution of heavy
elements.
Care should be taken when choosing the amplitude $C(r)$ to
avoid unphysically large azimuthal gravity forces near the origin. 
The reader is referred to Vorobyov \& Shchekinov (\cite{VS}) for a detailed discussion
on this subject.

\section{Code description}
\label{code}
An Eulerian finite-difference code is used to solve equations~(\ref{first})-(\ref{third})
in the polar coordinates ($r,\phi$). The basic algorithm of the code is similar
to that of the ZEUS code presented by Stone \& Norman (\cite{SN}). The operator
splitting is utilized to advance in time the dependent variables in two
coordinate directions. The advection is treated using the consistent
transport method of Stone and Norman and the van Leer interpolation  scheme.
The resolution is $400\times 400$ grid points and the radial size of the grid
cell is 50~pc. The code performs well on the angular momentum conservation problem
(Norman et al. \cite{Norman}). This test problem is essential for the adequate modelling of
rotating systems with radial mass transport.

\section{Initial configuration}
\label{init}
\begin{figure}
  \resizebox{\hsize}{!}{\includegraphics{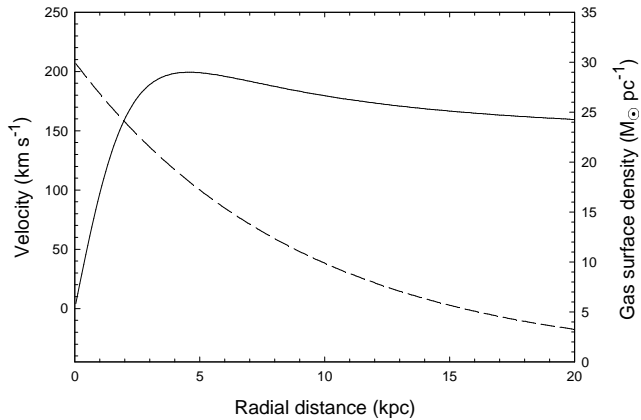}}
      \caption{The initial rotation curve (solid line) and gas surface
      density distribution (dashed line).}
         \label{fig1}
\end{figure}

The initial equilibrium gas disk configuration is initialized by fixing the
radial profile of the gas disk and calculating the rotation curve that balances the 
combined  gravity force of the dark matter halo and axisymmetric part of
the stellar disk. We assume that
the gas disk has an exponentially declining density profile 
\begin{equation}
\Sigma_{\rm g}=\Sigma_{\rm g0} \exp(-r/r_{\rm g}),
\end{equation}
with the central surface density $\Sigma_{\rm g0}=30~M_\odot$~pc$^{-2}$,
and radial scale length $r_{\rm g}=9$~kpc. The total mass of the gas disk
within the computational domain ($r=20$~kpc) is
$M_{\rm g}=1.0 \times 10^{10}~M_\odot$. Hence, the gas disk contains only
a small fraction ($\sim 10\%$) of the total mass in the computational domain
and, to a first approximation, we can neglect its self-gravity.
The initial gas density profile and the initial rotation curve are shown
in Fig.~\ref{fig1} by the dashed and solid lines, respectively.

Once the equilibrium gas disk is set, we slowly introduce the non-axisymmetric
part of the stellar disk. Specifically, $\Phi_{\rm s2}$ is multiplied by
a function $\epsilon(t)$, which has a value of $0$ at $t=0$ and linearly
grows to its maximum value of $1.0$ at $t\ge200$~Myr.
It takes a few hundred Myr for the gas disk to adjust to the spiral distortion
and develop a spiral structure. 

We do not attempt to follow the exact chemical evolution of our model galaxy.
Instead, we simply assume that at the time when spiral structure appears the
model galaxy has already developed the abundance of elements that is typical for disk galaxies. 
We further focus on the
purely dynamical influence of stellar spiral density waves 
on the distribution of pre-existing heavy elements
in the galactic disk. 
Taking the above considerations into account, we construct
the radial distribution of heavy elements by assuming that they
have a given abundance at corotation and choose an appropriate negative radial gradient.
The effect of star formation will be considered in a subsequent paper.

\section{Radial migrations of heavy elements at corotation}
\label{results}
\begin{figure}
  \resizebox{\hsize}{!}{\includegraphics{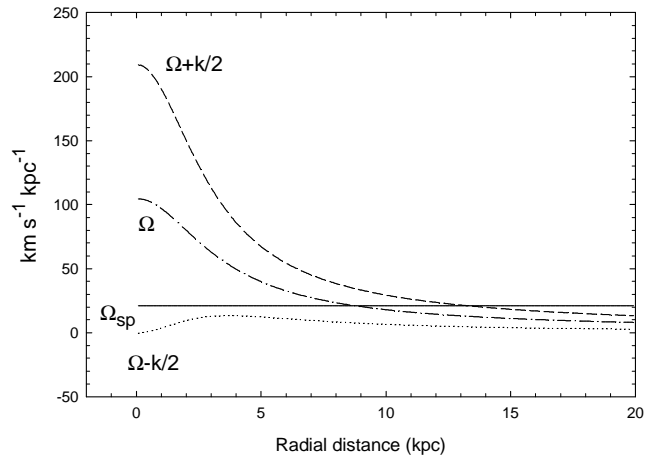}}
      \caption{Radial profiles of the gas angular velocity $\Omega$ (dotted-dashed
      line) and $\Omega \pm k/2$ (dashed and dotted lines, respectively), where $k$ is the
      epicyclic frequency. The angular velocity of spiral pattern $\Omega_{\rm
      sp}$ is shown by the solid line.}
         \label{fig2}
\end{figure}

As was demonstrated by Sellwood \& Binney (\cite{Sellwood}) and Vorobyov \&
Shchekinov (\cite{VS}), spiral stellar density waves at corotation are 
powerful drivers of radial migration of stars and dust.
In this section, we study the effect that spiral density waves
have on the radial abundance distribution of heavy elements in the warm
gas component of spiral galaxies.
Two models of spiral stellar density waves are considered.
In both models, the stellar density wave
(the gravitational potential of which is given by Eq.~[\ref{nonsym}]) 
rotates counter clockwise at an angular velocity  $\Omega_{\rm sp}=21$~km~s$^{-1}$~kpc$^{-1}$.
This choice of $\Omega_{\rm sp}$ places the corotation of the gas disk at $\approx 9$~kpc and the outer
Lindblad resonance at $\approx 13$~kpc as is shown in Fig.~\ref{fig2}. The inner Lindblad resonance is absent.
The pitch angle is $i=25^\circ$,
if not otherwise stated. According to Kennicutt (\cite{Ken1}), a pitch angle of $i=25^\circ$
is typical for Sc galaxies. The radial position of resonances 
may vary along the azimuth during the evolution of the gas disk by $\approx 5\%$.

In the first model (hereafter, model~1), the amplitude of the stellar
spiral gravitational potential $C(r)$ is chosen so that it is maximal near
the position of corotation. Consequently, the gas disk has a strong spiral
response to the underlying spiral stellar density wave at both sides of corotation. 
In the second model (hereafter, model~2),
the spiral stellar density wave is mostly localized within corotation.
Observationally, models~1 and 2 represent two types of spiral galaxies in which
corotation is situated approximately in the middle of the spiral pattern and at the very end of
it, respectively.

\begin{figure}
  \resizebox{\hsize}{!}{\includegraphics{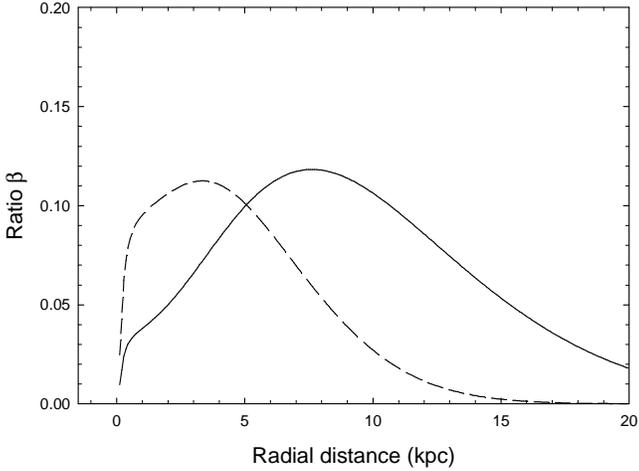}}
      \caption{The ratio $\beta$ of the maximum non-axisymmetric gravitational
      acceleration (at a given radial distance $r$) $|{\bl\nabla} \Phi_{\rm
      s2}|=[(\partial \Phi_{\rm s2}/\partial r)^2+(r^{-1}\partial \Phi_{\rm s2}/\partial
      \phi)^2]^{1/2}$ to the total axisymmetric gravitational acceleration
      $|(\partial \Phi_{\rm s1}/\partial r+ \partial\Phi_{\rm h}/\partial
      r)|$  as a function of radial distance in model~1 (solid line) and
      model~2 (dashed line).}
         \label{fig3}
\end{figure}

Since we do not consider the effect of star formation on the abundance
distribution of heavy elements, we choose not to focus on any particular
element like oxygen or iron but instead use an abundance of a  generic
heavy element X defined as
\begin{equation}
\left[{X\over H}\right]=log_{10}\left( {\Sigma_{\rm x}\over \Sigma_{\rm g}} \right) -  
log_{10} \left( {\Sigma_{\rm x}\over \Sigma_{\rm g}} \right)_{\rm cr}.
\label{abundance}
\end{equation}
The last term in equation~(\ref{abundance})
gives the abundance of heavy element X at corotation. 
In the following, we assume that element X has an initial radial abundance
gradient of $-0.05$~dex~kpc$^{-1}$. According to Daflon \& Cunha (\cite{Daflon}),
an average radial slope of such elements as C, N, O, Mg, Al, Si, and S in the
Galactic disk is $-0.042\pm0.007$~dex~kpc$^{-1}$. The oxygen abundance
in a sample of external galaxies studied by Zaritsky, Kennicutt, \&
Huchra (\cite{Zaritsky}) shows a wide spread in the radial slopes ranging
from $-0.009\pm 0.01$~dex~kpc$^{-1}$ for NGC~1365 to $-0.231\pm
0.022$~dex~kpc$^{-1}$ for NGC~3344 (some galaxies have a positive slope).
On average, most galaxies in their sample have radial oxygen gradients
near $-0.05$~dex~kpc$^{-1}$. Once the radial abundance gradient and the
abundance at corotation are fixed, we can use equation~(\ref{abundance})
to determine $\Sigma_{\rm x}$ and equations~(\ref{first})-(\ref{third})
to compute the dynamics of the gas and heavy element X. 

\subsection{Model~1}
\label{model1}
The amplitude $C(r)$ of the spiral stellar gravitational potential determines the response
of the gas and consequently the appearance of a spiral pattern in the gas disk.
We adopt the following expression $C(r)=C_0(r)^{\alpha(r)}$. Here, $C_0(r)$ is a linear 
function of $r$ which has a value of 0 at $r=0$~kpc 
and attains its maximum value of 0.0018 (in dimensionless units) at $r=20$~kpc.
The exponent $\alpha(r)$ decreases linearly
with radius from $\alpha=2$ at $r=0$~kpc to $\alpha=-0.1$ at $r=20$~kpc. 
The resulting ratio $\beta(r)$ of the maximum non-axisymmetric gravitational acceleration
$|{\bl\nabla} \Phi_{\rm s2}|$ (at a given radial distance $r$) 
to the total axisymmetric gravitational acceleration $|(\partial \Phi_{\rm s1}/\partial r+ \partial\Phi_{\rm h}/\partial
r)|$ is shown in Fig.~\ref{fig3} by the solid line. 
Since the non-axisymmetric gravitational acceleration scales
as $|{\bl\nabla} \Phi_{\rm s2}| \propto 1/r^{1-\alpha(r)}$, the ratio $\beta$ 
increases with radius at $r<7.5$~kpc, and decreases at $r\ga 7.5$~kpc. 
We note that the maximum, non-axisymmetric
perturbing force never exceeds $12\%$ of the total axisymmetric gravity
force.  At the position of corotation $r_{\rm cr}\approx 9$~kpc, $\beta$ 
approximately equals $11\%$.

\begin{figure}
  \resizebox{\hsize}{!}{\includegraphics{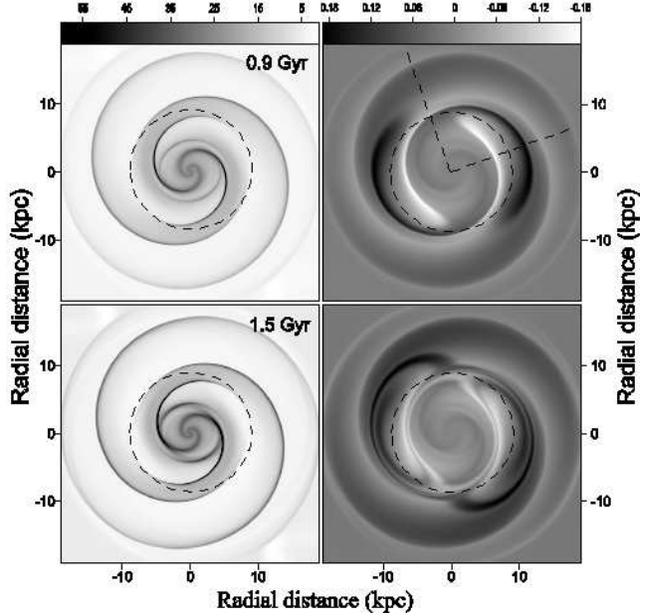}}
      \caption{Left panels -- the gas surface density distribution, right
      panels -- the distribution of residual abundances of element X at two different evolutionary
      times
      as indicated in the left panels. The approximate position of corotation
      is shown by the dashed circle. The left scale bar is in $M_\odot$~pc$^{-2}$
      and the right scale bar is in dex.}
         \label{fig4}
\end{figure}

Once the spiral stellar density wave is slowly turned on,
the initially axisymmetric gas disk responds to a disturbing gravitational field of stellar
spirals and develops a spiral  pattern.
The left panels of Fig.~\ref{fig4} show the evolution of the gas surface
density at two different times as indicated in each panel. The two-arm
spiral structure in the gas disk is clearly visible in these images.
A strong gas response to the gravitational field of the spiral density
wave is seen inside and outside corotation, the position of which is sketched 
by a dashed circle. Enhancements and depressions in the gas surface density
distribution by approximately a factor of 2  as compared to the unperturbed 
distribution are seen in the spiral arms and between them, respectively.
The gas response to the spiral stellar density wave is minimal around corotation.

\begin{figure}
  \resizebox{\hsize}{!}{\includegraphics{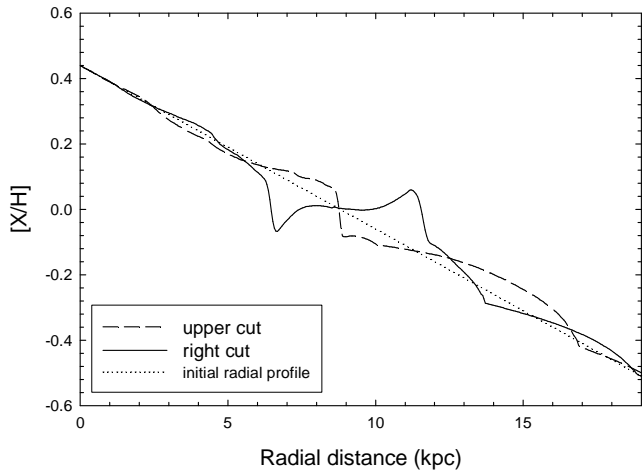}}
      \caption{The radial abundance distribution of element X determined
      at $t=0.9$~Gyr  along two radial cuts made through the gas disk. 
      The directions of the right and upper cuts are shown in the upper panel of  Fig.~\ref{fig4} by
      the dashed lines.
      The dotted line shows the initial axisymmetric abundance distribution
      of element X.   }
         \label{fig5}
\end{figure}

The right panels in Fig.~\ref{fig4} show the residual abundances [X/H]$_{\rm
res}$, which are obtained by subtracting the current abundances [X/H] at a given
position in the disk from those at $t=0$~Gyr. If there were no radial and/or
azimuthal migrations of element X due to spiral stellar density waves, the
residual abundance would be zero during the consequent evolution of the
gas disk. However, a considerable redistribution of
element X near corotation is evident in Fig.~\ref{fig4} as indicated by large
deviations of [X/H]$_{\rm res}$ from zero. 
The maximum deviations of $\pm
0.18$~dex are found between the spiral arms, whereas the deviation is noticeably smaller 
in the region where spiral arms cross corotation. 
This implies that the radial profiles of [X/H] {\it may be different along different
radial directions}. Indeed,  Fig.~\ref{fig5} shows the radial distribution of
[X/H] along two radial cuts that are plotted by the dashed lines
in the upper-right panel of Fig.~\ref{fig4}. 
The first cut (hereafter, the upper cut) is made through the point in the disk where the gas spiral arms cross the corotation circle.
The second cut (hereafter, the right cut) is made at a $90^\circ$ angle to the upper cut.
The initial axisymmetric abundance distribution of element X is plotted
by the dotted line for comparison.
An obvious (and impressive) difference is seen between the radial [X/H] profiles along
the upper and right cuts. The former shows a sharp drop near the
position of corotation at $r\approx 9$~kpc, while the latter has a plateau
(with a size of approximately 4~kpc) at the position of corotation.
The radial profiles of [X/H] inside corotation
demonstrate substantially smaller deviations
from the initial axisymmetric radial profile, which implies little redistribution
of heavy elements in the inner galactic regions.

\begin{figure}
  \resizebox{\hsize}{!}{\includegraphics{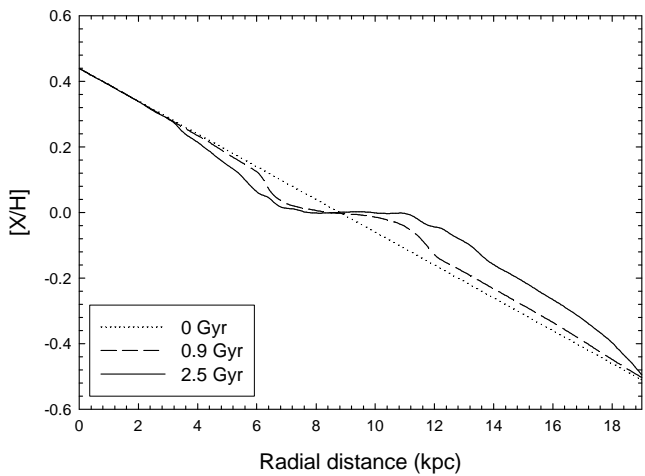}}
      \caption{Azimuthally averaged abundance distribution of element X
      at three different times as indicated in the legend. The development
      of a plateau on both sides of corotation ($\approx 9$~kpc) is evident.}
         \label{fig6}
\end{figure}

A small number of metallicity indicators (such as HII regions or planetary
nebula) usually makes it difficult 
to determine the radial abundance profiles along a particular radial direction in a galaxy.
In this respect, the azimuthally averaged radial abundance distribution
is of particular interest. The azimuthally averaged radial distribution of [X/H] is 
shown in Fig.~\ref{fig6} at two different times after the beginning of
simulations. The dotted line plots the initial axisymmetric abundance distribution
of element X.  
The development of a plateau in the azimuthally averaged profiles of [X/H] is evident
in Fig.~\ref{fig6}. 
Taking into account that the spiral perturbing force saturates after 200~Myr
from the beginning of simulations, only two revolution periods ($\sim 640$~Myr)
are needed to develop a significant plateau with a size of 3~kpc.
The plateau grows in time and reaches a size of approximately
4.5~kpc after 2.5~Gyr. The radial redistribution of element X with respect
to hydrogen is also visible in the plots -- a substantial amount of element
X (approximately $7\%$ by mass) is pushed
radially away in the disk by the action of spiral stellar density waves.

\begin{figure}
  \resizebox{\hsize}{!}{\includegraphics{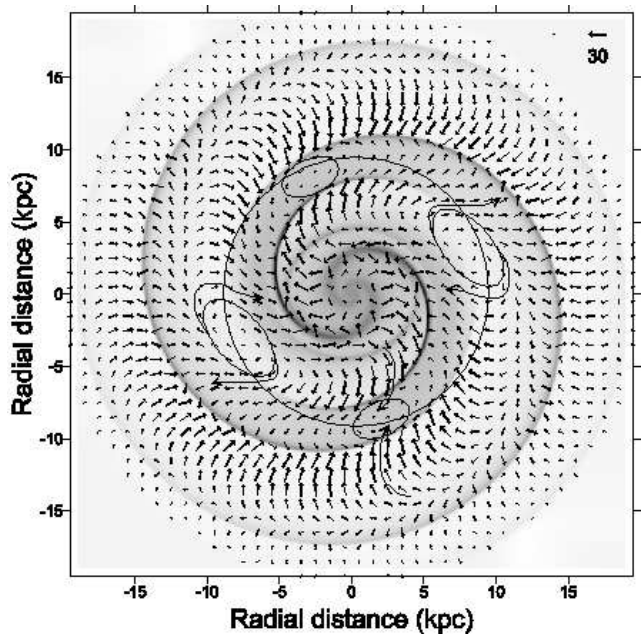}}
      \caption{The residual gas velocity field superimposed onto the gas surface
      density distribution in model~1 at $t=0.9$~Gyr. The locations of cyclones and anticyclones
      are shown with the smaller and larger ellipses, respectively. Large
      arrows sketch the gas flow in cyclones and anticyclones. Corotation
      is drawn by the circle. The scale vector is in km~s$^{-1}$. }
         \label{fig7}
\end{figure}

As was mentioned in the Introduction, the gravitational field of the stellar
spiral arms is known to radially mix the stars and cold molecular clouds and transport dust radially away in the disks of
spiral galaxies (Sellwood \& Binney \cite{Sellwood}, Vorobyov \& Shchekinov \cite{VS}).
In order to understand the effect that the stellar spiral arms have on the
distribution of heavy elements in the warm gas phase of spiral galaxies, we consider
a large-scale gas flow in the disk of our model galaxy. Figure~\ref{fig7}
shows the residual velocity field of gas superimposed on the gas surface density distribution at $t=0.9$~Gyr.
The residual velocity field of the gas is obtained by subtracting the
circular motion of gas due to the combined axisymmetric gravitational potential
of stars and dark matter halo from the total gas velocity. 
The residual velocity field shows a complicated flow
pattern. However, several regular features of the gas flow are clearly
visible in Fig.~\ref{fig7}. On average, the gas streams radially inward along the outer
edge of a spiral arm and radially outward along the inner edge. The radial
velocities are considerable and may become as high as $\pm (25-30)$~km~s$^{-1}$ near
the corotation circle.
These large-scale radial gas motions are caused by the gravitational field of stellar spiral
density waves. A hydrogen atom (as well as any other element coupled to hydrogen) is
exposed to an additional gravitational attraction of the stellar density wave when it approaches the spiral arm.
If the atom is closer to the concave part of the spiral, it will be pushed
radially outward. 
On the other hand, if the atom is closer to the convex part of the spiral,
it will be given an inward radial pull. This two-fold action of the spiral stellar
density wave on the dynamics of a single particle is schematically depicted 
by  Vorobyov \& Shchekinov (\cite{VS}) in their fig.~6. The gravitational
field of stellar spirals 
can cause a considerable change in the angular momentum of a particular element
(it could be either a gas parcel or a star) but the the total angular momentum
of the system remains largely unchanged.

The effect of spiral stellar density waves on the dynamics of gas
(and heavy element admixtures within it) is strongest around the corotation resonance,
where a gas parcel {\it rotates in phase with the stellar spiral density wave}
and is subject to the maximal radial migrations. Considerably away from corotation, the gas
parcel frequently passes through the spiral arms and the gravitational influence of the latter
mostly cancels out (as is seen in the trajectories of test particles in
Fig.~5 of Vorobyov \& Shchekinov \cite{VS}). 
The net result of the radial migrations of heavy elements 
is a flattening of the pre-existing negative radial abundance gradients near corotation.
Since the abundance of element X is larger inside corotation than outside
it, the radial gas flow exports many more atoms
of element X outside corotation than it takes from there. This process
continues until a near homogeneous abundance distribution is set on both sides
of corotation.

The smearing of negative abundance gradients near corotation can
be better understood in terms of anticyclones that develop between the spiral
arms and are sketched in Fig.~\ref{fig7} by the larger ellipses.
The anticyclone flow of gas is approximately shown by the long arrows.
It is clearly seen that the anticyclones push the gas from inside corotation
to the regions outside it and vice versa, thus producing an effective
mixing of elements and flattening of negative abundance gradients 
at the position of anticyclones (i.e. between the spiral arms).
The small ellipses show the location of two cyclones that form 
in the disk near the point where spiral arms cross the corotation circle.
Two approaching streams of gas in the cyclone (shown by the long arrows in
Fig.~\ref{fig7}) meet near corotation, which
creates a contact discontinuity and a subsequent sharp drop in the 
abundance distribution at corotation. Such a step-like feature
is clearly seen in the radial abundance distribution of element
X calculated along the radial cut that passes through the position
of the cyclone (see the dashed line in Fig.~\ref{fig5}). 

The formation of cyclones and anticyclones in the gas disk is a consequence
of a non-axisymmetric gravitational field of the stellar spiral density wave.
A gas parcel that orbits sufficiently close to corotation
($\la 2$~kpc on both sides of it) is exposed to a strong resonant forcing of
the spiral gravitational field. Such resonant forcing produces anticyclones
and associated considerable radial migrations of gas and heavy elements.
An example of trajectories around corotation is shown by the solid line
in Fig.~5 of Vorobyov \& Shchekinov (\cite{VS}). 
The same type of ``horse-shoe'' orbit was found by Sellwood
\& Binney (\cite{Sellwood}) in a stellar disk around corotation.
We note that the existence of such cyclone and anticyclone motions 
was indeed revealed in the gas disks of NGC~3631 (Fridman et al. \cite{Fridman1}) and NGC~157
(Fridman et al. \cite{Fridman2}). 
The anticyclone motions of cold gas clouds around corotation in spiral galaxies 
were also suggested by Sellwood \& Preto (\cite{Sellwood2}).

Another important implication seen in Fig~\ref{fig7} is that the gas {\it
is not pushed away from corotation everywhere along the corotation perimeter}. 
An assumption of the outflow of gas away from corotation usually involves
a two-fold effect of spiral shocks (triggered by stellar density
waves) on the gas that passes through the shock front.  Because of differential
rotation, the gas that orbits inside/outside corotation is decelerated/accelerated at the shock
front and is driven radially inward/outward from corotation.
This conclusion may be misleading because it completely neglects the long-range
influence of the gravitational field of spiral arms,
which appear to have a much stronger effect on the gas dynamics than the spiral
shock fronts. The gas is indeed pushed away from corotation by
the action of anticyclones but is returned to corotation by the action of cyclones.
However, the azimuthally averaged gas surface density distribution 
shows a mild gas depression near corotation
as compared to the initial distribution, implying that anticyclones are more powerful than cyclones.

\begin{figure}
  \resizebox{\hsize}{!}{\includegraphics{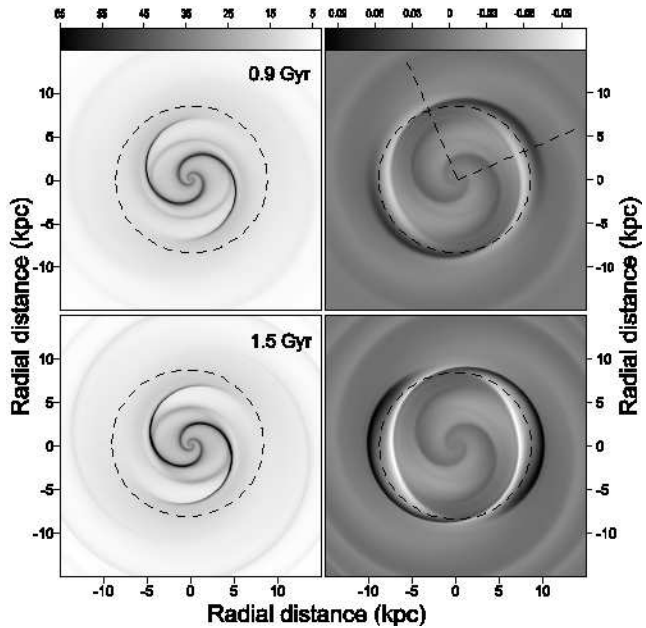}}
      \caption{The same as Figure~\ref{fig4} but for model~2.}
         \label{fig8}
\end{figure}

\subsection{Model~2}
Simulations of the previous section have shown that the spiral stellar
density waves induce anticyclone flows of gas near corotation and, by
doing so, flatten the pre-existing negative abundance gradients of heavy
elements in the warm gas. This effect is limited to several
kiloparsecs on each side of corotation, and the abundance distribution
retains (on average) its initial slope at radial distances that are considerably far from corotation.
This implies that if the stellar spirals are entirely confined within corotation,
we should not expect to see a considerable flattening of the negative abundance gradients.
We check this assumption by considering model~2 in which 
we choose the amplitude $C(r)$ of the spiral stellar gravitational potential 
so that the stellar density waves are mostly localized within corotation.
This means that they have a negligible effect on the gas dynamics outside corotation.
The same expression for the amplitude $C(r)$ as in model~1 is adopted. However,
the exponent $\alpha(r)$ decreases linearly
with radius from $\alpha=2$ at $r=0$~kpc to $\alpha=-2.4$ (in
contrast to $\alpha=-0.1$ in model~1) at $r=20$~kpc.
The resulting profile of the maximum non-axisymmetric gravitational acceleration
to the total axisymmetric gravitational acceleration $\beta(r)$ is shown in Fig.~\ref{fig3}
by the dashed line.  The ratio $\beta$ 
increases with radius at $r<4$~kpc, and decreases at $r\ga 4$~kpc. 
We note that the maximum, non-axisymmetric
gravitational acceleration never exceeds $11\%$ of the total axisymmetric gravity
force.  At the position of corotation $r_{\rm cr}\approx 9$~kpc, $\beta$ 
approximately equals $4\%$ and it sharply decreases beyond corotation. 

\begin{figure}
  \resizebox{\hsize}{!}{\includegraphics{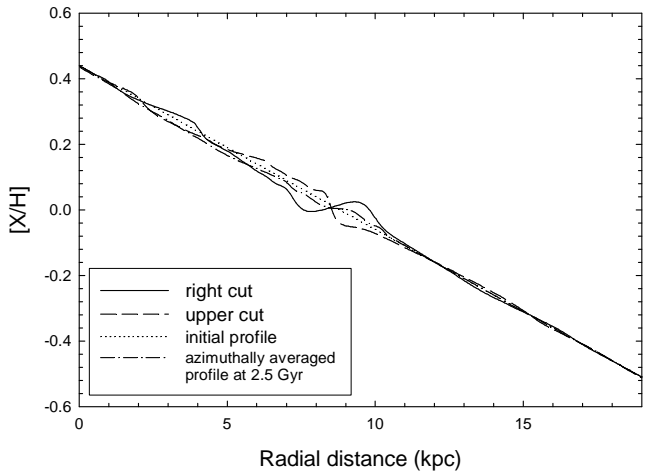}}
      \caption{The radial abundance distribution of element X determined
      at $t=0.9$~Gyr  along two radial cuts made through the gas disk. 
      The directions of the right and upper cuts are shown in the upper panel of  Fig.~\ref{fig8} by
      the dashed lines.
      The dotted line shows the initial axisymmetric abundance distribution
      of element X.  The dashed-dotted line show the azimuthally averaged
      abundance distribution at $t=0.9$~Gyr. }
         \label{fig9}
\end{figure}

The left panels in Fig.~\ref{fig8} show the evolution of the gas surface
density at two different times as indicated in each panel.
The gas response to the gravitational field of spiral stellar density waves
in model~2 is remarkably different from that of model~1. The gas spiral
pattern in model~2 is completely localized within corotation, which is drawn in 
Fig.~\ref{fig8} by a dashed circle. A weak spiral
structure can be traced in the gas disk beyond corotation but its amplitude is negligible
compared to that inside corotation.
The right panels in Fig.~\ref{fig8} show the distribution of residual abundance 
[X/H]$_{\rm res}$. The maximum deviations from the
initial abundance distribution of element X are found near corotation. However,
they are (at least) a factor of 2 smaller than in model~1. 
As in the previous section, we make two radial cuts in the gas disk at
$t=0.9$~Gyr and calculate the abundance profiles of element X along these
cuts. The cuts are drawn in the upper-left panel of Fig.~\ref{fig8} by the dashed lines. 
The directions of the cuts are the same as in model~1, which allows for
a direct comparison between both models.
The solid and dashed lines in Fig.~\ref{fig9} show the resulting
distributions of [X/H] obtained along the upper and right cuts, respectively. The dotted line
shows the initial [X/H] distribution for comparison.
It is obvious that the deviation from the initial unperturbed axisymmetric
distribution  is much smaller in model~2 than in model~1. The abundance
profiles do have a small plateau and a mild discontinuity at the position
of corotation, but this effect is mostly cancelled out after azimuthal
averaging. As a result,  the azimuthally averaged abundance distribution
of element X plotted in Fig.~\ref{fig9} by the dashed-dotted line is almost indistinguishable from
the initial unperturbed distribution.

\begin{figure}
  \resizebox{\hsize}{!}{\includegraphics{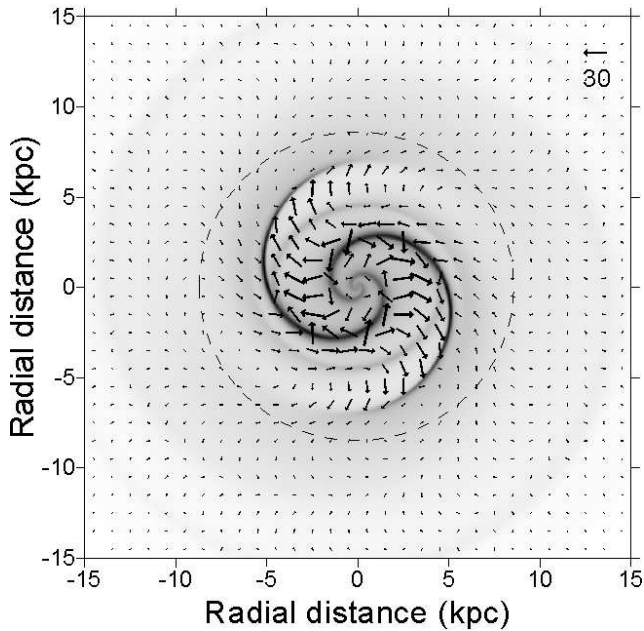}}
      \caption{The residual gas velocity field superimposed onto the gas surface
      density distribution in model~2 at $t=0.9$~Gyr. Corotation is dran by the dashed
      circle.  The scale vector is in km~s$^{-1}$. 
      Cyclones or anticyclones are hardly visible in the residual gas velocity
      field. }
         \label{fig10}
\end{figure}

The difference between the two models in the efficiency of flattening of 
element abundance gradients can be readily understood
if we consider the residual gas velocity field which is obtained by subtracting the
circular velocity of gas (due to the axisymmetric gravitational field) from the total gas
velocity.
Figure~\ref{fig10} presents the residual gas velocity field superimposed
onto the gas surface density distribution at $t=0.9$~Gyr. The approximate
position of corotation is drawn by the dashed line. 
Similarities and differences in the overall gas flow in model~1 and model~2 are evident.
In both models, the gas streams inward/outward along the outer/inner edges of spiral arms.
However, in model~2 the radial velocities around corotation are at least
an order of magnitude smaller ($\approx 2-3$~km~s$^{-1}$) than in model~1.
Very weak cyclones and anticyclones are seen around corotation.
Inside corotation ($r<7$~kpc), considerable radial velocities up to $35$~km~s$^{-1}$
are evident in Fig.~\ref{fig10}. Nevertheless, these radial flows involve
little radial transport of heavy elements (and consequent flattening
of negative abundance gradients), because gas particles frequently cross
the spiral arms and effectively cancel out the gravitational influence
of the latter.
As a result, gas particles and heavy element admixtures move around the galactic centre
on closed low-eccentricity circular orbits, as shown by the dashed line in Fig.~5 of Vorobyov
\& Shchekinov (\cite{VS}).

\section{Flattening of abundance gradients along the Hubble sequence}
\label{hubble}
The tightness of the spiral pattern defined by the pitch angle of spiral
arms is one of the fundamental criteria in Hubble's (\cite{hbl}) classification
scheme of spiral galaxies. Kennicutt (\cite{Ken1}) has confirmed that there
exist a correlation between the pitch angle and the Hubble type of spirals. 
A smooth, monotonic increase in the openness of spiral arms with
later Hubble type is detected on the average, but large variations in the pitch angle 
within a given type are still present. 

\begin{figure}
  \resizebox{\hsize}{!}{\includegraphics{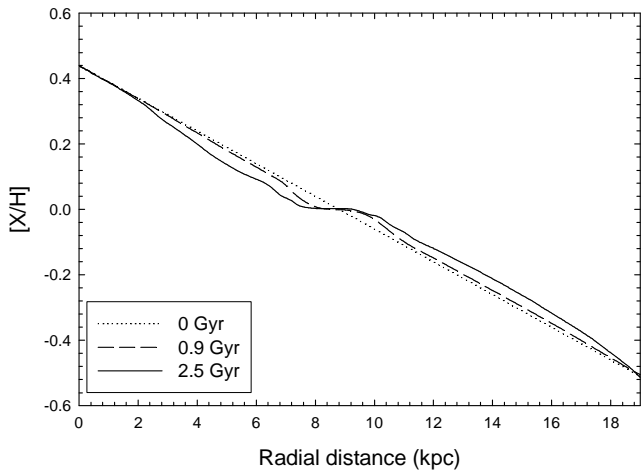}}
      \caption{Azimuthally averaged abundance distribution of element X
      at three different times as indicated in the legend.} 
         \label{fig11}
\end{figure}

\begin{figure}
  \resizebox{\hsize}{!}{\includegraphics{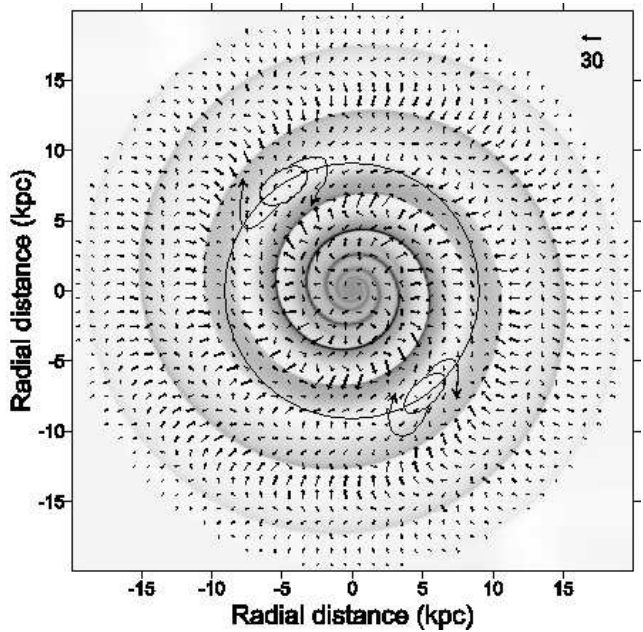}}
      \caption{The residual gas velocity field superimposed onto the gas surface
      density distribution in the model with $i=10^\circ$ at t=0.9~Gyr. 
      The locations of anticyclones are shown with the ellipses. Large
      arrows sketch the gas flow in anticyclones. Corotation
      is drawn by the circle. The scale vector is in km~s$^{-1}$. } 
         \label{fig12}
\end{figure}

In the previous section, we studied open spirals with a pitch angle of $i=25^\circ$. 
According to Kennicutt (\cite{Ken1}), this value is near the upper limit
($i\approx30^\circ$) for the late-type Sc galaxies.
In this section, we consider more tightly-wound spirals 
typical for the early-type Sa-Sb galaxies.
The model galaxy has the same initial parameters as model~1 of Sect.~\ref{model1}
but a pitch angle $i=10^{\circ}$.
The resulting azimuthally averaged radial abundance profiles of element
X are plotted in Fig.~\ref{fig11} at two different evolutionary times. 
The initial radial profile is shown by the dotted line for comparison.
It is evident that the size of the plateau around corotation decreases by at least a factor
of 2 as compared to that in model~1. This effect can be understood if
we consider the residual gas velocity field shown in Fig.~\ref{fig12} for
the model with $i=10^{\circ}$.  This velocity field is calculated in the
same manner as in Sect.~\ref{model1} and is superimposed on the gas surface density
distribution at $t=0.9$~Gyr. The position of corotation is drawn by the circle and 
the anticyclones are drawn by the ellipses.
A comparison of Fig.~\ref{fig12} and Fig.~\ref{fig7} indicates that the anticyclone flows of gas are much 
less pronounced in tightly-wound galaxies with $i=10^\circ$ than in galaxies
with $i=25^\circ$. The radial gas velocities and
the overall extent of anticyclones  are noticeably smaller in tightly-wound spirals.
For instance, the maximum  and average radial velocities around corotation in
the $i=25^\circ$ model~1 are $\pm(25-30)$~km~s$^{-1}$ and $\pm(5-8)$~km~s$^{-1}$, respectively.
The inward radial velocities are usually larger then the outward ones by
approximately $10\%-20\%$. On the other hand, 
in the  $i=10^\circ$  model with more tightly-wound arms the maximum and average radial velocities
are $\pm(5-8)$~km~s$^{-1}$ and $\pm(1-2)$~km~s$^{-1}$, respectively. It is not surprising
that the efficiency of radial redistribution of heavy elements  and the sizes
of plateaus around corotation are noticeably lower in  the $i=10^\circ$ model than in the $i=25^\circ$
model~1.

The radial positions of a logarithmic stellar spiral at zero phase ($\phi=0$, see
eq.~\ref{nonsym}) can be determined as
\begin{equation}
r_{\rm n}=r_{\rm sp} \exp ({2 \pi n \tan(i)/ m})
\label{spiral}
\end{equation}
where $n=0, \pm 1, \pm 2, \cdots$  and $r_0=r_{\rm sp}$ is the characteristic radius of 
the stellar spiral at $phi=0$.
Equation~\ref{spiral} becomes inapplicable when the separation between
{\it the gas spirals} is considered, because there exist
a non-negligible azimuthal phase shift between the stellar and gas spirals caused
by the non-solid-body rotation curve of the gas disk.
This phase shift is particularly large outside corotation where the gas
rotation curve declines with radius. As a consequence, the separation between
the gas spirals decreases considerably, especially in the case of larger pitch angles. 
Therefore, we use a visual estimate for the radial separation
between the gas spirals.

Our numerical simulations indicate that 
the anticyclone flows  {\it are bounded by the adjacent gas spiral arms}.
This implies that the sizes of plateaus in the radial profiles of heavy elements
should be proportional to the radial
separation between the gas spiral arms on both sides of corotation. 
Indeed, a visual inspection of Fig.~\ref{fig7} and Fig.~\ref{fig12} indicates
that the radial separation is at least a factor of 2 larger in the $i=25^\circ$ 
model than in the $i=10^\circ$ model. Consequently, the size of the plateau
around corotation is also approximately a factor of 2 larger in the $i=25^\circ$ model. 
Hence, the radial separation between the gas spiral and an associated anticyclone
activity is expected to increase along the Hubble sequence.
It is now not surprising that both Grand-Design galaxies NGC~3631 and NGC~157, 
where giant anticyclones have been observed (Fridman et al. \cite{Fridman1,Fridman2}), are
of Sc and Sbc types, respectively.  
We conclude that plateaus in the radial abundance distribution of heavy
elements have a tendency to increase in size along the Hubble sequence.

Two-armed spiral galaxies have been considered so far.
In multi-armed galaxies ($m>2$), the radial separation between the gas spiral
arms is smaller than in two-armed galaxies with the same pitch angle. 
This should also reduce the sizes of anticyclones and the associated
efficiency of radial heavy element redistribution. We leave this issue for a subsequent study.

\section{Summary}
\label{sum}
We have studied numerically the dynamics of warm gas and heavy elements in spiral galaxies.
Two types of two-armed spiral galaxies are considered in which
corotation is situated approximately in the middle of the spiral pattern (type~1) and at the very end of
it (type~2), respectively. Type~2 spiral galaxies show little radial redistribution of
heavy elements in the warm gas disk. 
Conversely, type~1 spirals  are distinguished by a substantial radial redistribution
of heavy elements around corotation.

Strong cyclones and anticyclones around corotation
are generated by the non-axisymmetric gravitational field
of spiral stellar density waves in type~1 galaxies. The anticyclone flows transport gas
from inside corotation to  the regions outside it and vice versa. 
If the radial abundances of heavy elements are characterized by a negative
slope, the anticyclones bring many more atoms of heavy elements 
outside corotation than they import inside corotation.
This results in a flattening of radial abundance profiles at
the position of anticyclones after two revolution periods of a galaxy. 
On the other hand, the cyclone flows generate in-going and out-going streams
of gas along the spiral arms. These streams meet at corotation, producing a contact discontinuity in
the gas flow and associated step-like radial profiles of heavy element
abundances at the position of cyclones. Nevertheless, the azimuthally averaged
radial abundance distributions of heavy elements show {\it  a well-defined plateau
on both sides of corotation}, implying that anticyclones are more powerful
in transporting the heavy elements than cyclones.

The sizes of plateaus around corotation in the azimuthally averaged radial abundance distributions
of heavy elements
are expected to increase along the Hubble sequence in spiral galaxies with equal
number of arms.
Our numerical simulations show that in two-armed Sa-Sb galaxies with
a pitch angle of $10^\circ$, the plateau has a maximum size of 2~kpc. 
In contrast, Sc galaxies with a pitch angle of $25^\circ$ have 
the plateau that can become as large as 4.5~kpc.
A growing efficiency of radial mixing of heavy elements (and an associated
flattening of negative abundance gradients) 
along the Hubble sequence
is related to the increasing radial separation between the spiral arms
and a consequent increase in the strength of anticyclones.

A considerable portion of total gas mass in a spiral galaxy may be in the form 
of cold molecular hydrogen clouds. Our numerical simulations show that the efficiency 
of radial mixing grows as the gas temperature drops. A decrease in the restoring 
force of pressure gradients is responsible for this effect. However, our numerical 
hydrodynamics code is not appropriate for the modelling of molecular cloud dynamics, 
for which the sticky particle codes are well 
suited. Sellwood \& Binney (\cite{Sellwood}) and Sellwood \& Preto (\cite{Sellwood2}) 
have employed N-body simulations to study the dynamics of cold molecular clouds. They 
have reported the development of anticyclonic motions at corotation 
and predicted the flattening of any metallicity gradients within the 
disc.

In this paper, we neglect the effect of continuous star formation. It 
is a computationally difficult task to self-consistently include the 
production of heavy elements in the hydrodynamics code. A possible exception is 
oxygen which is released mostly by short-living massive stars, for 
which the instantaneous recycling approximation can be used (e.g. 
Acharova et al. \cite{Acharova}). Our preliminary numerical 
simulations indicate that the on-going star formation in the spiral 
arms may change the shape of the plateau at corotation. More 
specifically, a mild minimum and maximum in the azimuthally averaged 
radial abundance profile of oxygen may develop at the inner and outer 
sides of corotation, respectively. Similar shapes in the radial 
abundance distribution of oxygen near corotation were reported by 
Acharova et al. (\cite{Acharova}). The results of this study will  be 
presented in a follow-up paper.

\section*{Acknowledgments}
The author is grateful to the referee, J. R. D. L\'{e}pine, for useful suggestions that 
helped to improve the final presentation.
The author is also thankful to Shantanu Basu for help with English language usage
and Yu. A. Shchekinov for valuable comments. Support from a CITA National Fellowship
is gratefully acknowledged.

\end{document}